\documentstyle[preprint,aps]{revtex}
\preprint{P-93-07-065}
\begin{document}
\draft
\title{{\bf
Quantum Disordered Regime and Spin Gap \\
in the Cuprate Superconductors}}
\author{\large V. Barzykin$^{(1)}$, D. Pines$^{(1)}$,
A. Sokol$^{(1,2,4)}$, \\ and D. Thelen$^{(1,3)}$}
\address{$^{(1)}$Department of Physics,
$^{(2)}$Materials Research Laboratory, \\
and $^{(3)}$Science and Technology Center for Superconductivity, \\
University of Illinois at Urbana-Champaign, Urbana, IL 61801 \\
$^{(4)}$L.D.\ Landau Institute for Theoretical Physics, Moscow, Russia}
\maketitle
\begin{abstract}
We discuss the crossover from the quantum critical, $z\!=\!1$, to
the quantum disordered regime in high-T$_c$ materials
in relation to the experimental data on the nuclear relaxation,
bulk susceptibility, and inelastic neutron scattering.
In our scenario, the spin excitations develop a gap
$\Delta\!\sim\!1/\xi$ well above
T$_c$, which is supplemented by the quasiparticle gap below T$_c$.
The above experiments yield consistent estimates for the
value of the spin gap, which increases as the correlation length
decreases.
\end{abstract}
\pacs{PACS: 74.65.+n, 75.40.Cx, 75.40.Gb, 76.60.-k}

\narrowtext
The underdoped cuprates YBa$_2$Cu$_3$O$_{6.63}$,
YBa$_2$Cu$_4$O$_8$, and La$_{1.85}$Sr$_{0.15}$CuO$_4$,
display strikingly
different magnetic behavior from that measured for the fully doped system,
YBa$_2$Cu$_3$O$_7$. In YBa$_2$Cu$_3$O$_7$, the uniform spin susceptibility,
$\chi_0$, is essentially temperature independent, and the product
$^{63}T_1T$ of the
nuclear spin-lattice relaxation time and the temperature has monotonic
temperature dependence; in the underdoped compounds $\chi_0$ is temperature
dependent  while the temperature dependence  of
$^{63}T_1T$ is non-monotonic.
Recent measurements of the spin-echo decay time,  $T_{\rm 2G}$\
\cite{Pennington:Slichter},
for YBa$_2$Cu$_3$O$_{6.63}$\ \cite{TakigawaT2},  together with the earlier
measurements for YBa$_2$Cu$_3$O$_{6.9}$ \cite{Imai7},
provide additional valuable
information on such {\em spin gap} phenomena in the underdoped systems.

Alternative physical origins of the spin gap have been proposed recently by
Millis and Monien \cite{Millis:Monien}
and Sokol and Pines \cite{Sokol:Pines},
hereafter SP. Millis and Monien \cite{Millis:Monien} argued on the basis of
bulk susceptibility measurements
that the spin gaps in  YBa$_2$Cu$_3$O$_{6.63}$\
and  La$_{1.85}$Sr$_{0.15}$CuO$_4$\ possess a different physical origin,  with
the gap in YBa$_2$Cu$_3$O$_{6.63}$\ arising from a magnetic coupling between
adjacent CuO$_2$ layers,  while for La$_{1.85}$Sr$_{0.15}$CuO$_4$ it is
attributed to the spin density wave fluctuations of a metal. On the other hand,
SP applied scaling  arguments to the analysis of the NMR experiments to
determine the magnetic phase  diagram of the cuprates. They  argued  that
at high
temperatures these materials, and the closely related system,
YBa$_2$Cu$_4$O$_8$,  are in the quantum critical (QC), $\rm z=1$, regime
(see \cite{CHN,Chubukov:Sachdev}),
characterized by a temperature
independent ratio $\rm ^{63}T_1T/T_{2G}$,  and a linear
dependence on T of $\rm ^{63}T_1T$
\cite{Chakravarty:Orbach,Chubukov:Sachdev},
while the common physical origin of the spin gap
behavior  these systems exhibit at lower temperatures
is the suppression of the
spectral weight for frequencies,  $\rm \omega\!<\!\Delta\!\sim\!c/\xi$,
characteristic of the quantum disordered (QD) regime.
In the quantum critical regime \cite{CHN,Chubukov:Sachdev},
$ \bar{\omega}\!\sim\!\xi^{-1}\!\sim\!T $ and $z\!=\!1$,
while in the quantum disordered
regime $\xi$ saturates ($\xi$ is the antiferromagnetic
correlation length, $\bar{\omega}$
characteristic energy scale, and $z$ dynamical exponent).
In the present
communication  we show that the spin susceptibility and neutron scattering
measurements on these underdoped cuprates are consistent  with  the SP magnetic
phase diagram, reconcile the low frequency  spin dynamics measured by NMR with
the results of neutron scattering experiments,  and give estimates of the spin
gap, $\Delta$,  the spin wave velocity, c, and $\xi$.

We consider first the Knight shift and bulk susceptibility measurements. As
shown in Ref.\cite{Chubukov:Sachdev}, in the QC, $z=1$,  regime, $\chi_0$ is
linear in $T$, in agreement with numerical calculations \cite{Glenister:Singh}
as well as  experiment for La$_2$CuO$_4$. In the renormalized classical regime,
$\chi_0$ is finite at $T\!=\!0$; on the other hand, in the QD regime, the
susceptibility rapidly (exponentially for the insulator) decreases due to the
presence of the gap in the spin excitation spectrum,  and tends to zero as
$T\!\to\!0$ \cite{Chubukov:Sachdev}.  As may be seen in Fig.\ref{chi}, at high
temperatures  $\chi_0(T)$ displays the expected linear in T behavior  for
La$_2$CuO$_4$, La$_{1.85}$Sr$_{0.15}$CuO$_4$, $\rm YBa_2Cu_3O_{6.63}$, and
YBa$_2$Cu$_4$O$_8$ with somewhat varying slope.

The crossover from the QC to the QD regime is clearly visible  in the
experimental  data shown in Fig.\ref{chi}, apart from the insulator.  As
discussed in SP, the temperature $\rm T^*$, at which $\rm ^{63}T_1T$ starts to
deviate from its linear behavior ($\rm T^* \sim 200K$ for $\rm
YBa_2Cu_3O_{6.63}$ and $\rm YBa_2Cu_4O_8$, and $\rm T^* \sim 125K$ for $\rm
La_{1.85}Sr_{0.15}CuO_4$), signals the beginning of this crossover. This
downturn, when $\chi_0$ rapidly  (although not exponentially  in the doped
case)
decreases for $T\!\lesssim\!\Delta$, reflects the suppression of  the low
frequency spin excitations.  As may be seen in Fig.\ref{chi},   the downturn in
$\chi_0(T)$ begins
at nearly the same temperature  as that at  which $^{63}T_1T$
begins to deviate from its linear behavior. It follows that,  as expected for
the QD regime, the spin gap influences the low frequency dynamics
both near ${\bf q}\!=\!0$ and ${\bf q}\!=\!(\pi,\pi)$.
The experimental  finding
that the $\rm ^{17}O$ spin lattice relaxation rate in $\rm YBa_2Cu_3O_{6.63}$
\cite{TakigawaT1}  and $\rm YBa_2Cu_4O_8$ \cite{Zheng}
follows the temperature  dependence  of
$\rm \chi_0(T)$ $\rm [(^{17}T_1T)^{-1}\!\sim\!\chi_0(T)]$
is thus consistent with our scenario.
If one assumes, with SP, that the effect of the quasiparticles on
the QNL$\sigma$ model parameters is somewhat less for
$\rm La_{1.85}Sr_{0.15}CuO_4$ than in $\rm YBa_2Cu_3O_{6.63}$, and ascribes
this to a smaller hole density in the former material, and also,
as Dupree {\em
et al.} \cite {Dupree} argue,  that $\rm YBa_2Cu_4O_8$ possesses a hole density
comparable to that found in $\rm YBa_2Cu_3O_{6.8}$, we see that the temperature
at which the downturn occurs increases with hole concentration,
going from $\rm
\sim 100K$ for $\rm La_{1.85}Sr_{0.15}CuO_4$ to
$\rm \sim 240K$ for $\rm YBa_2Cu_3O_{6.63}$ and
$\rm YBa_2Cu_4O_8$.  Moreover, since
the magnitude of $\rm \chi_0(T)$ in the QC and QD regime may be expected to
increase with hole doping, this trend would be
likewise evident in Fig.\ref{chi},
once one
assigns a hole density in $\rm La_{1.85}Sr_{0.15}CuO_4$ similar to that
found for $\rm YBa_2Cu_3O_{6.55}$,  and takes its orbital Knight shift to be
in the range 0.02\% to 0.076\%.

A close examination of the data shows that {\em two}  successive downturns are
visible in Fig.\ref{chi}, with only the second  corresponding to
the superconducting transition (Fig.\ref{chi}).
This supports the view that in the
superconducting state of QD materials, two separate  gaps are present; a charge
gap which reflects the pairing of  quasiparticles in the superconducting state,
and  a spin gap $\Delta\!\sim\!1/\xi$,
which reflects the QD behavior modified, in the
superconducting state, by the pairing correlations.

We turn next to
the evidence for a spin gap and a cross-over from  the QC to the
QD regime found in the neutron scattering experiments on the
YBa$_2$Cu$_3$O$_{6+x}$\ system.  Consider first  YBa$_2$Cu$_3$O$_{6.6}$, where
the measurements of Ref.\cite{Shirane}  show that the q-integrated
(i.e.\ local)
dynamical response function,   $\chi''_L(\omega) = \smallint d^2{\bf q} \,
\chi''({\bf q},\omega)$,  is a universal function of $\omega/T$ \cite{Shirane}
which behaves as  $\chi''_L\!\sim\!\omega/T$
for high temperatures (i.e.\ small
$\omega/T$).
This behavior is consistent with the QC scaling found  for
this compound in the analysis of the NMR experiments  \cite{Sokol:Pines}.
Moreover, the value of $\bar{\chi}_L(\bar{\omega})=T \chi_L(\omega)$ inferred
from experiment \cite{Sato} is in good agreement with the universal function
calculated using the $1/N$ expansion \cite{Chubukov:Sachdev}. For low
temperatures, the suppression of $\chi''_L$ for YBa$_2$Cu$_3$O$_{6.63}$,  which
is seen as an abrupt deviation  from the universal behavior of $\chi''_L$  as a
function of $\omega/T$, is an indication of the spin gap phenomenon.  From an
analysis of their experimental results, Tranquada {\em et al.}
\cite{Shirane} conclude
that the magnitude of the spin gap for this material is  $\rm \Delta \simeq
10\mbox{meV}$.

Since at $T\!=\!0$,
$\chi''_L(\omega)\!=\!0$ for $\omega\!<\!\Delta$, an alternative
approach is to associate the maximum of $d\chi''_L(\omega)/d\omega$  with the
spin gap magnitude, $\Delta$.  Rossat-Mignod {\em et al}.\
\cite{Rossat-Mignod:1} find that for YBa$_2$Cu$_3$O$_{6.69}$ this maximum
occurs at 16 meV and we take this to be the value of $\Delta$ for this
compound.  At lower hole doping, in the range $0.4\!<\!x\!<\!0.5$
just above the
metal-insulator transition at $\rm x \cong 0.41$,  according to
Ref.\cite{Sokol:Pines}, YBa$_2$Cu$_3$O$_{6+x}$ is expected to be in the QC
region for $T \geq 50\mbox{K}$. While no spin gap was seen in the
non-superconducting  YBaCuO$_{6.4}$,
Rossat-Mignod {\em et al.}
\cite{Rossat-Mignod:1} have reported the existence of a small gap in
the low temperature superconducting region of YBaCuO$_{6.45}$ and
YBaCuO$_{6.51}$.

At higher hole doping, in the heavily-doped YBa$_2$Cu$_3$O$_{6+x}$ with $x
\simeq 1$, a value of the spin gap $\sim 26\mbox{meV}$  has been reported by
Rossat-Mignod {\em et al.}
using  unpolarized neutrons \cite{Rossat-Mignod:1}.
Mook {\em et al.} \cite{Mook}
have used polarized neutrons  to study YBa$_2$Cu$_3$O$_7$ and have
reported that  a low-frequency suppression of the signal, ascribed
to the spin gap formation, may be present at $\rm \sim 35\mbox{meV}$; they
find as well a magnetic peak at 41meV.  Inelastic neutron experiments in the
intermediate region of  oxygen concentrations  6.75 and 6.9 have also been
reported in Ref.\cite{Sato}, but the measurements have been
performed only at low
temperatures, where the superconducting gap,
which complicates such an analysis
and may affect the spin gap, is already developed.
We show in Fig.\ref{gaps} the values
of $\Delta$ which have been reported for various oxygen concentrations.

Our conclusion that the spin gap increases as the doping increases is somewhat
unexpected in view of previous analyses  \cite{Millis:z=2},  but may be easily
understood in terms of the QNL$\sigma$ model, where $ \Delta=c/\xi$, provided
$\xi$ increases faster than $\rm c$  as the doping decreases.  One can,
moreover, combine the results of neutron scattering and NMR experiments to
obtain an estimate of the doping dependence of c and $\rm \xi$.

We consider first $\rm YBa_2Cu_3O_{6.63}$, for which we estimate $\rm \Delta
\cong 12\mbox{meV}$ by interpolating between the experimental results for $\rm
YBa_2Cu_3O_{6.6}$ and $\rm YBa_2Cu_3O_{6.69}$.
This value is somewhat less than
that found by applying scaling arguments to the experimental results for $\rm
T_{2G}(T)$, shown in Fig.\ref{T2G}.
Applying the approach of \cite{Chubukov:Sachdev},
one obtains that in the QC regime,
$T_{2G}(T)$ is linear in T, while the ratio of this
slope and its extrapolated intercept
is the product of a universal number, $\zeta$, and
$\Delta$.  On taking $\zeta$
as obtained from the leading order of a $\rm 1/N$
expansion, one finds
$\rm \Delta = 18\mbox{meV}$; the discrepancy between the
two results plausibly reflects next-order corrections to $\zeta$,
as well as
uncertainties in the determination of $\Delta$ in a neutron scattering
experiment.

A different approach is to make use of the scaling result
$\chi_Q\!\sim\!\xi^{2-\eta}$, $\eta\!\ll\!1$,
and neglect any possible temperature dependence of $\alpha$ in
$\rm \chi_Q\!=\!\alpha\xi^2$.
To the extent this holds,
the full temperature dependence of $\rm \xi(T)$ may be determined
directly from $T_{2G}$, as shown in Fig.\ref{T2G}.
Moreover, to the extent that the QNL$\sigma$ model is valid,
despite the absence
of a rigorous Lorentz invariance of the action, the relation $\rm \Delta =
c/\xi$ should still approximately apply in the QD regime at low
temperatures.  With these assumptions, for a given choice of $\alpha$,
$\rm \chi_Q$, $\rm \xi$, and c may be estimated by
comparing the neutron scattering result
$\rm \Delta \cong 12\mbox{meV}$ with
the NMR results for $\rm ^{63}T_1$ and $\rm T_{2G}$ and
assuming Lorentzian form of $\chi(\widetilde q)$.
For convenience, we introduce a dimensionless quantity
$\rm \bar{\alpha}_{TP}=\alpha/\alpha_{TP}$,
where $\rm\alpha_{TP}\!=\!14.8states/eV$ is a value of $\alpha$ estimated
by Thelen and Pines \cite{Thelen:Pines} for $\rm YBa_2Cu_3O_{6.9}$.
We find that at low temperatures, in the QD regime,
$\rm \xi\!\simeq\! 4.2 \bar{\alpha}_{TP}^{-1}$ and
$\rm c\!\simeq\! \bar\alpha_{TP}^{-1} \cdot 0.19eV\AA$,
while in the QC regime,
one has $\rm \xi^{-1}= \bar{\alpha}_{TP}
\cdot 0.1 (1\!+\!T/100K)$. We note in passing that T-linear QC behavior
of both $\xi^{-1}$ and $T_{\rm 2G}$
in a doped antiferromagnet has been
recently reported by Glenister, Singh, and one of us (A.S.)
\cite{Sokol:Glenister:Singh}; see also Ref.\cite{Chubukov:Sachdev}.

We examine next the extent
to which the neutron scattering experimental results
on $\rm La_{1.85}Sr_{0.15}CuO_4$ are consistent with the NMR results, the
magnetic phase diagram of SP,  and our proposed variation of $\Delta$ with
hole
concentration.  Here there are two issues.  First, if the hole doping in $\rm
La_{1.85}Sr_{0.15}CuO_4$ is comparable
to that found in $\rm YBa_2Cu_3O_{6.55}$,
one would expect  from Fig.\ref{gaps}  that the spin gap $\Delta$ for $\rm
La_{1.85}Sr_{0.15}CuO_4$ is $\rm \sim 6\mbox{meV}$.  While the experiments
reported in Ref.\cite{Mason}
show no clear evidence for such a gap,
we find an upturn near $\rm \omega \leq 6\mbox{meV}$ in
$\rm \chi_L''(\omega)$ at $\rm 35K$ measured by
Thurston {\em et al.} \cite{Shirane:La},
as might be expected if $\rm \Delta \sim
6\mbox{meV}$ for this material.  Second, while the incommensurate spin
fluctuation peaks, at ($\rm \pi,\pi\pm\delta$) and
($\rm \pi\pm\delta,\pi$), where $\rm \delta \!=\!0.245\pi$,
reported in Ref.\cite{Mason},
are not incompatible with our picture,
such a large value of $\delta$
is incompatible with NMR experiments.
The results of Ref.\cite{Sokol:Gagliano:Bacci} confirm the prediction
\cite{Imai:Slichter}
that the hyperfine constants for the 214 systems are quite close to
those inferred for the 123 systems and specifically, for $\rm
YBa_2Cu_3O_{6.63}$ \cite{Monien:Pines:Takigawa}.
On taking the neutron results for $\rm \chi''({\bf q},\omega)$
it is straightforward to show that at
$\rm 100\mbox{K}$, the copper-to-oxygen
relaxation rate ratio,
$\rm (^{17}T_1/^{63}T_{1,\parallel})\!\lesssim\!35$
while the $\rm ^{63}Cu$ relaxation rate anisotropy ratio,
$(^{63} T_{1,\parallel}/^{63}T_{1,\perp})\!\sim\!4.5$.
These results, which reflect the leakage of correlations
away from $(\pi,\pi)$,
brought about by the comparatively large value of $\delta$, are to be
compared to those found experimentally:
$\rm (^{17}T_{1,\parallel}/^{63}T_{1,\parallel})\!\gtrsim\!80$
\cite{Reven} and
$\rm (^{63}T_{1,\parallel}/^{63}T_{1,\perp})\!\sim\!2.6$.

Can the NMR and neutron scattering results be reconciled?
The problem is not with the
magnitude of the measured spin fluctuation peaks.
The value of $\rm ^{63}T_{1,\parallel}T$
is $\rm \sim 5.5 \times 10^{-2}K\cdot sec$
at $\rm T\!=\!35\mbox{K}$,
when calculated from the neutron scattering results of
Ref.\cite{Mason} for q-integrated intensity
$\rm \chi_L(\omega)$,
assuming commensurability.
This compares favorably with the NMR result,
$\rm ^{63}T_{1,\parallel}T\!\sim\!5.9\cdot10^{-2}K\cdot sec$;
indeed the
slightly larger value at $\rm 35\mbox{K}$ may reflect the continued upturn
associated with a spin gap $\rm \sim 6\mbox{meV}$. The problem rather lies in
the assumption that the peaks are incommensurate.  If instead, as suggested
independently by Slichter \cite{Slichter} and Phillips\cite{Phillips} what is
being observed is discommensuration, associated with the
formation of domains in
the Cu-O plane, the NMR and neutron scattering experiments
can be reconciled, as
shown in recent calculations by Monthoux\cite{Monthoux}.  Thus in neutron
scattering one is likely seeing both commensurate spin fluctuation peaks and
domain structure.  Two other experimental results lend support to this
proposal:  the appearance of a slightly incommensurate peak in $\rm
YBa_2Cu_3O_{6.6}$ \cite{Shirane}  which might be expected to show behavior not
very different from that of $\rm La_{1.85}Sr_{0.15}CuO_4$,
and the appearance of
a second $\rm ^{63}Cu$ resonance line in both the Sr-doped $\rm 214$ systems
\cite{Yoshimura} and $\rm La_2CuO_{4.032}$ \cite{Hammel},
which would plausibly be
associated with $\rm ^{63}Cu$ nuclei located in or near domain walls.

In summary,
we have shown that NMR, bulk susceptibility, and neutron  scattering
data in the underdoped materials are consistent
with the picture of Sokol and
Pines \cite{Sokol:Pines}, where the spin gap onset is associated with the
crossover from the  QC, $z=1$, regime to
the low temperature QD regime. The spin
gap magnitudes in YBa$_2$Cu$_3$O$_{6.63}$, YBa$_2$Cu$_4$O$_8$, and
La$_{1.85}$Sr$_{0.15}$CuO$_4$\ compounds determined
from different  experiments
are consistent with each other. The magnitude
of the spin gap gradually {\em
increases} as doping increases; since the quasiparticle
contribution also increases, this eventually  leads to the smearing of the gap
and  a crossover to the overdamped, $z\!=\!2$, regime in the  fully doped
materials. Such a doping dependence is consistent with  the theoretical
prediction that $\Delta\!\sim\!c/\xi$, provided the  correlation length
decreases more rapidly then c as the doping increases.

We are grateful to P.C.\ Hammel,
T.\ Imai, T.E.\ Mason, A.J.\ Millis, R.R.P.\ Singh,
C.P.\ Slichter, and M.\ Takigawa for many stimulating discussions,
to T.\ Imai, C.P.\ Slichter, and M.\ Takigawa for communicating  their
experimental data prior to publication, and to G.\ Aeppli and T.E.\ Mason
for cooperation in making comparisons between the NMR and neutron
scattering data.
This work was supported by the NSF Grant DMR89-20538
through the Materials Research Laboratory and DMR91-20000 through
the Science and Technology Center for Superconductivity.

\begin{figure}
\caption{The spin susceptibility for:
$\Box$ - La$_2$CuO$_4$
\protect\cite{Johnston},
from the bulk susceptibility;
\protect\rule{8pt}{8pt} -
La$_{1.85}$Sr$_{0.15}$CuO$_4$
\protect\cite{Ishida},
from Cu Knight shift;
{\protect\Large$\bullet$} - YBa$_2$Cu$_3$O$_8$
\protect\cite{Machi},
from Cu Knight shift;
{\protect\Large$\circ$} ($^{63}$K$_{ab}$), $\triangle$ ($^{17}$K$_{ax}$),
$\nabla$ ($^{17}$K$_{c}$), $\Diamond$
($^{17}$K$_{iso}$) - YBa$_2$Cu$_3$O$_{6.63}$
\protect\cite{TakigawaT1},
from the Knight shift.
Solid lines show linear fits to the high temperature parts of
the respective data, except for $\rm YBa_2Cu_3O_{6.63}$, where
the slope could not be determined because little high temperature
data is available; it has been chosen to be the same as
for $\rm YBa_2Cu_4O_8$ instead.}
\label{chi}
\end{figure}

\begin{figure}
\caption{The values of the spin gap in
YBa$_2$Cu$_3$O$_y$ at different oxygen concentrations $y$
determined in inelastic neutron scattering experiments,
see the text for discussion; \protect\rule{8pt}{8pt} -
M. Sato {\em et al.} \protect\cite{Sato},
$\Diamond$ - J. Rossat-Mignod {\em et al}
\protect\cite{Rossat-Mignod:1},
$\Delta$ - J. M. Tranquada {\em et al} \protect\cite{Shirane},
{\protect\Large$\circ$} - H. A. Mook {\em et al} \protect\cite{Mook}.}
\label{gaps}
\end{figure}

\begin{figure}
\caption{The Gaussian spin-echo decay time for YBa$_2$Cu$_3$O$_{6.63}$,
after Ref.\protect\cite{TakigawaT2}.
The solid line is the high temperature linear fit (QC regime);
the dotted line represents its saturation in the QD regime.
An alternate axis shows the
temperature dependence of $\xi^{-1}$ as given by the scaling relation
$T_{\rm 2G}\!\sim\!1/\xi$.}
\label{T2G}
\end{figure}

\end{document}